\documentclass[a4paper,fleqn,usenatbib]{mnras}

\usepackage{newtxtext,newtxmath}

\usepackage[T1]{fontenc}
\usepackage{ae,aecompl}

\usepackage{graphicx}	
\usepackage{amsmath}	
\usepackage{amssymb}	

\title[M 42 structure functions]{An analysis of the turbulence in the central region of M 42 through structure functions}

\author[Anorve-Zeferino GA]{
	G.A. Anorve-Zeferino,$^{1}$\thanks{E-mail: Alex.Anorve@protonmail.com}
	\\
	$^{1}$\'Ecole Polytechnique, Route de Saclay, Palaiseau, 91128, France
}

\date{Accepted XXX. Received YYY; in original form ZZZ}

\pubyear{2018}

\begin{document}
\label{firstpage}
\pagerange{\pageref{firstpage}--\pageref{lastpage}}
\maketitle

\begin{abstract}
Here, we analyse the character of the turbulence of the Huygens Region in the Orion Nebula (M 42) using structure functions. We compute the second order  structure function of a high resolution velocity map in H$\alpha$  obtained through the {\emph MUSE} instrument. Ours is one of the few works  that follows a mathematically sound methodology for calculating the second order structure function of astronomical velocity fields.  Because of that our results will be useful for future comparisons with other studies of M 42 or other regions.  We first analyse the Probability Distribution Function (PDF) and found it consistent with  those resulting from numerical simulations of solenoidal turbulence. After a further analysis of the data, we found two possible separate motions or at least regimes in the region. This is confirmed later through the calculation of several filtered structure functions. We found that the turbulence in the Huygens Region is between the Kolmogorov regime ($S_2\propto\delta r^{2/3}$) and the Burgers regime  ($S_2\propto\delta r$). We found that the turbulence in the region consists on two flow regimes that reproduce a generalised Larson's Law, $S_2\sim\delta r^{0.74-0.76}$.
\end{abstract}




\section{Introduction}

The current standard view  is that turbulence is ubiquitous in H II regions, molecular clouds and star forming regions in general; see \citet{Klessen},   \citet{Glover}, \citet{Tremblin}, \citet{Mellema} and references therein.  The internal motions of the gas that comprise them are dominated by supersonic turbulence, the radiation field of young massive stars and also magnetic fields trapped within the ISM, see  \citet{Gritschneder} and \citet{FederrathK}. Supersonic turbulence can be driven by powerful superwinds [\citet{Anorve-Zeferino1} and \citet{Anorve-Zeferino2}],  the feedback from UV radiation, magnetic pressure,  magnetic rotational instability, galactic shear, SNe explosions,  gravity/accretion, cloud-cloud collisions, spiral-arm compression and some others, all of which can generate turbulence which can provide important  support against  gravitational collapse in molecular clouds and H II regions and also control their star formation, see \citet{Hartquist} and \citet{Federrath2017}.

Nevertheless, the velocity field of H II regions and molecular clouds can only be observed  in projection since we only have line-of-sight  information that can be synthesized in position-position-velocity (PPV) cubes  constructed after the detection and analysis of emission lines. Nevertheless, this provides the opportunity of knowing at least the line of sight motion inside these regions and try to extract information through diverse methods in order to unveil the physics behind it. One of such methods is the use of (transversal) $S_{\rm p}$-structure functions to try to diagnose turbulence. In this Paper, we obtain the second order structure function of the Huygens Region in the Orion Nebula using the centroid velocities of its H$\alpha$ emission lines.

The p-th order transversal $S_{\rm p}$-structure function is given by\footnote{ see~\citet{Federrath2009} and~\citet{Federrath2010} for examples of longitudinal structure functions of density and velocity fields from numerical simulations of compressive and solenoidal turbulence}

\begin{equation}
S_{\rm p}({\delta r})= \langle[u({{\vec{\bf r}}}+{\vec{ \delta\bf r}}) - u({{\vec{\bf r}}})]^p\rangle, \label{Eq:1}
\end{equation}

\noindent where the braces $<>$ indicate "average over the ensemble of realisations",  $u$ is the line-of-sight velocity and $\vec{\bf r}$ is the transversal radial vector. For the case of the PPV cubes here analysed, $u$ corresponds to the expansion velocity derived from  H$\alpha$ lines. For (quasi-)static fields, $<>$   is a type of spatial averaging.  At the time of calculating structure functions, authors calculate $<>$ through Fourier space methods, real space methods, or, instead, they use Monte Carlo algorithms like \citet{Konstandin} and more recently ~\citet{Boneberg}. However, the results from Monte Carlo calculations always stem from a partial use of the available data, which can produce severe errors. As we demonstrate here,  Monte Carlo methods  may not yield nor approximate correctly   second order structure functions.\footnote{ In the case of velocity fields  with nice properties like isotropy, homogenity or with symmetric simple PDFs it may be possible to obtain convergent structure functions but not in the general case because of the "mean field" effect discussed in Sections~\ref{data} and \ref{discussion}}  We directly prove this  by simply comparing the second order structure function obtained from a real space method and Fourier space methods --which lead to correlation functions, see \citet{Schulz-Dubois}--  with the results of a Monte Carlo algorithm.

 When normalised by $2\sum_{i\in\mathcal{L}} u_{i}^{2}$, with $\mathcal{L}$ being the  finite spatial extension of the velocity field $u$, the normalised second order structure function $S_{2,n}( {\delta r})\rightarrow 1$ for large enough ${\delta  r}$; in other words, the transversal  $S_{2}$    has a  well definite upper bound for large $\delta r$ when the data have a finite extension.\footnote{ in what follows we assume that $S_2$ is always normalised, so we will omit the subindex n}  Taking this into account and by using robust Fourier and real space algorithms to analyse the Huygens region of M 42, we will be (up to our knowledge) the first to compute correctly the $S_{2}$ corresponding to its H$\alpha$ emission and make the corresponding analysis. The PPV cubes we analyse were obtained by \citet{Weilbacher}.
  
 The paper is organized as follows. In Section 2 we analyse the  probability distribution function of the velocity field contained in the PPV cubes and select different thresholds for the minimum and maximum velocity to be considered. The previous allows to separate the image of the region in different segments of interest. In Section 3 we present the corresponding second-order structure functions. In Section 4 we discuss our results. The conclusions are presented in Section 5.
 
 \section{The data}\label{data}
 
 M 42 or the Orion Nebula, is one of the best and most studied Galactic H II regions and also the
 closest at a distance of approximately 440-pc. Because of that, it serves as a prototype for
 comparison with other H II regions. Its central region exhibits vigorous recent star formation
 and contains the densest nearby cluster of OB stars. The Nebula is located in front of the
 parent molecular cloud OMC-1, which makes it accessible for detailed study at all available
 wavelengths. Its ionized gas has been extensively studied spectroscopically in H$_\alpha$  and the radio bands e.g.~\citet{ODell3,vanderWerf}. Because of the great quantity of studies done about it, the Orion Nebula has become central for our understanding of massive star formation and stellar feedback.

\begin{figure} 
 	\centering    
 	\includegraphics[width=0.5\textwidth]{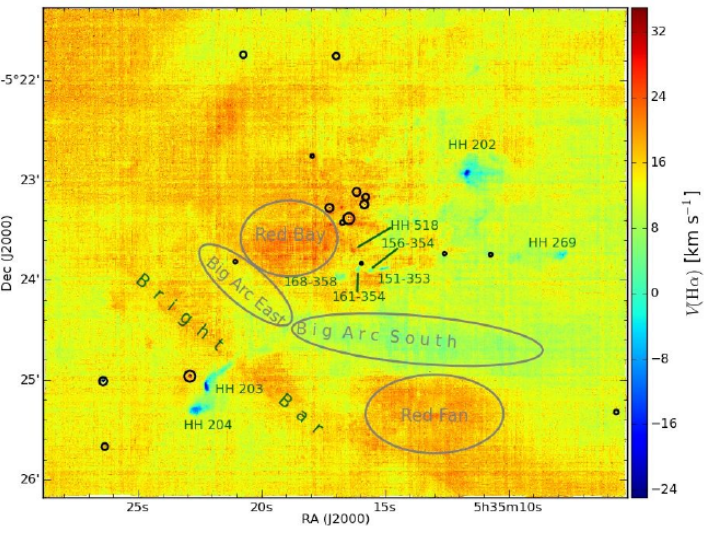}
 	\caption[ON]{The Orion Nebula velocity field in H$\alpha$. The ellipsoids indicate the zones where the brightest stars reside. Image from~\citet{Weilbacher}. }\label{Fig:fig1}
 \end{figure}

 Recently, the ionised gas in the central Huygens region of M 42  (with an angular size
 of $\sim 5.9' \times 4.9' \approx 0.76 \times 0.63$ pc${^2}$) was surveyed by~\citet{Weilbacher} in the wavelength range of $4595-9366$ \AA\,  using integral field spectroscopy. They obtained PPV maps in H$\alpha$ (see Figure~\ref{Fig:fig1})  using the \emph{MUSE} instrument at the ESO {\emph VLT}. Their spatial resolution was 0.2''.  They also used the [N II] and [S III] lines to derive the mean electron temperature and found $T_{e}\approx$  8500-9200 K.  Outside the ionization front they found electron densities in the range $N_e\approx$  500-10000 cm$^{-3}$. They also found that the layer of the ionization front has a higher electron density, $N_{\rm e}\approx$ 25000 cm${^{-3}}$.
 
 Using \emph{MUSE}, \citet{Weilbacher}  were able to obtain 2D maps of the ionized gas velocity from Gaussian centroids of the H$\alpha$ and other emission lines. In H$\alpha$, they found that the strongest features in velocity space are close to Herbig-Haro objects present in the region, some stars and also in a few broad zones labelled as the Bright Bar, Red Fan, Red Bay, Big Arc East and Big Arc South, see Figure~\ref{Fig:fig1}. The image size of the H$\alpha$ map in pixels is 1766 $\times$ 1476. The length corresponding to one pixel is 0.00043-pc.  In the map, the Herbig-Haro objects are marked with the label "HH".
 
\begin{figure}
	\centering    
	\includegraphics[width=0.5\textwidth]{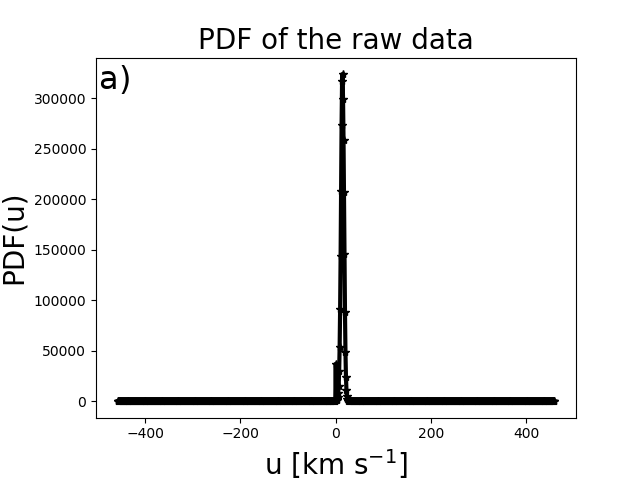}
	\includegraphics[width=0.5\textwidth]{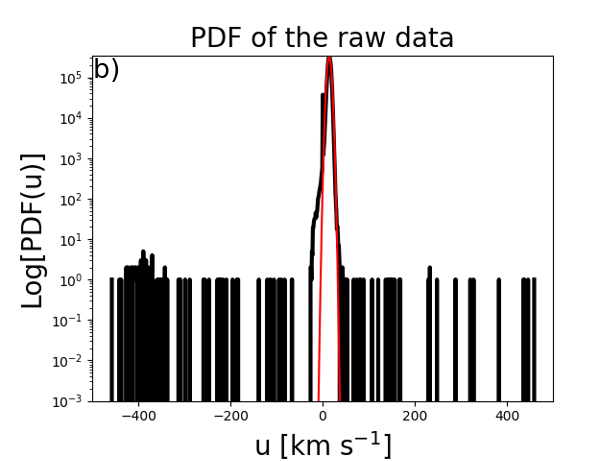}
	\includegraphics[width=0.5\textwidth]{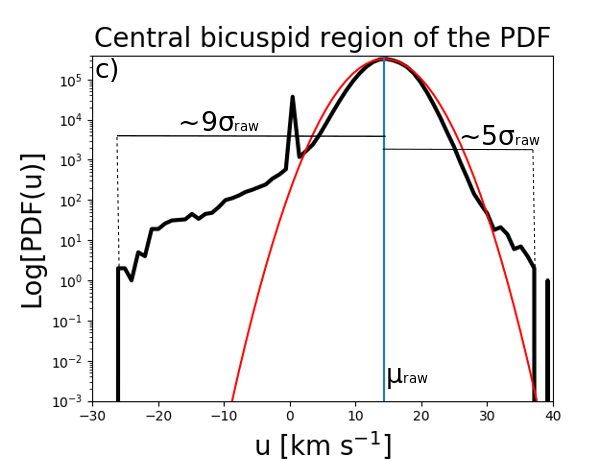}
	\caption[HAPDF]{a) PDF of the raw velocity data in H$\alpha$; b) semi-logarithmic plot of the previous PDF. The red line represents a Gaussian used as a reference fitting for the central part of the main cusp. The Gaussian has the same mean and standard deviation than the data within $-5\sigma/5\sigma$ from the raw data mean. The amplitude of the Gaussian was set to that of the tallest cusp; c) Zoom-in of the central bicuspid part of the PDF.  }
	\label{Fig:fig2}
\end{figure}

We obtained the un-normalised PDF (Probabilistic Distribution Function) of the raw H$\alpha$ data using 900 bins in order to make the secondary cusp to the left of the main central peak visible. The PDF is shown on Figure \ref{Fig:fig2}a. One can observe a very cuspy central distribution with velocities between $u\approx 1-23$ km s$^{-1}$, a small
cusp between $v\approx -1-1$ km s${^{-1}}$ and broad heavy tails that expand up to $u \approx -400-400$ km s${^{-1}}$. Broad heavy tails are usually found in studies of the PDF of interstellar turbulence, see~\citet{Federrath2010} and references therein. The mean of the data is $\mu_{u} = 14.38$ km s${^{-1}}$ and its standard deviation is $\sigma_u = 4.51$ km s${^{-1}}$ .

 Figures~\ref{Fig:fig2}b and~\ref{Fig:fig2}c show that the tails of the PDF are very heavy in comparison with those of the Gaussian fitted to the positive part of the central bicuspid region. The tails at the sides of this central region are discontinuous because they correspond to very small areas (from 1 to less than 10 pixels) with high positive or negative velocities. These very small areas are close to prominent features like Herbig-Haro objects and young stars, e.g. the small encircled region to the Northwest of HH 203. On the other hand, all negative velocities in the central bicuspid region correspond to small areas close to Herbig-Haro objects like HH 202, HH 203, HH 204 and HH 269. The areas of the regions with negative velocities  associated to the negative side of the central bicuspid region are larger than those associated to the discontinuous broad tails surrounding it.
	
It is noticeable that the left tail of the central bicuspid region is broader than  the right one. This asymmetry  has also been observed in PDFs obtained from numerical simulations of solenoidal and compressive turbulence, see figure A1 in~\citet{Federrath2013}. However,  differences between Figure~\ref{Fig:fig2}c and  figure A1 in~\citet{Federrath2013} are that, comparatively, the left tail in Figure~\ref{Fig:fig2}c is significantly broader than the right tail, it decreases  much less rapidly and it has a secondary cusp. Nevertheless,  except by these differences, the overall profile of the central bicuspid region of the PDF is consistent with that presented in figure A1 in \citet{Federrath2013}, in particular with the curve corresponding to solenoidal turbulence.

 The broad central sides of the central bicuspid region are important in determining the exponent of the inertial range of the second order structure function, much more important than the discontinuous tails at the sides, see Section~\ref{SF}. Thus, because of the long tails of the PDF, the data need to be filtered out in order to better understand the distribution around the two central cusps (in linear scale)  and its relation with the observed spatial distribution of velocities. We consider the next four cases.

\noindent {\bf{Filter 1}}

\begin{figure}
	\centering    
	\includegraphics[width=0.35\textwidth]{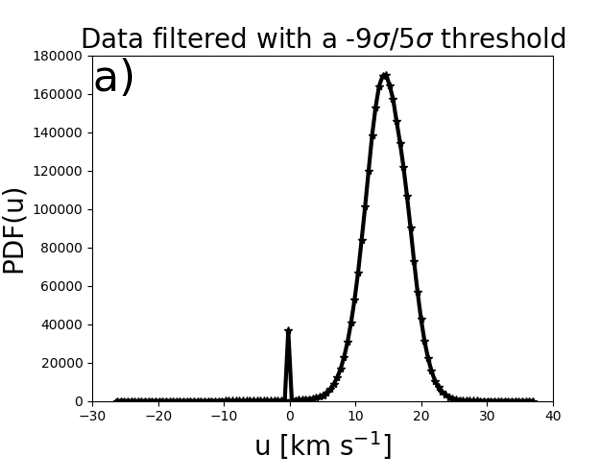}
	\includegraphics[width=0.35\textwidth]{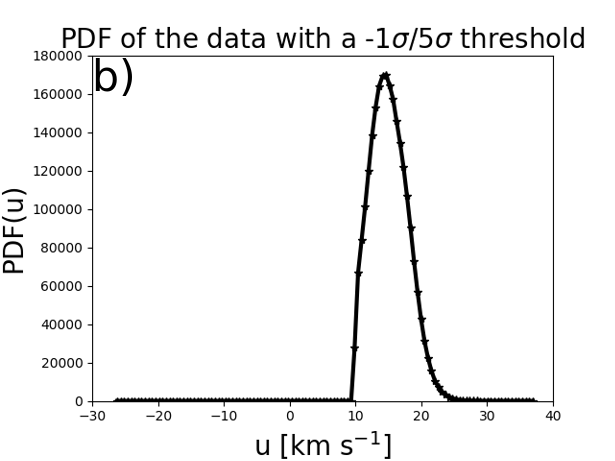}
	\includegraphics[width=0.35\textwidth]{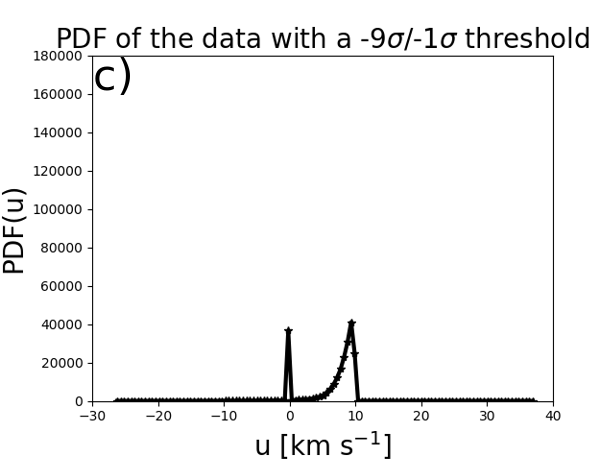}
	\includegraphics[width=0.35\textwidth]{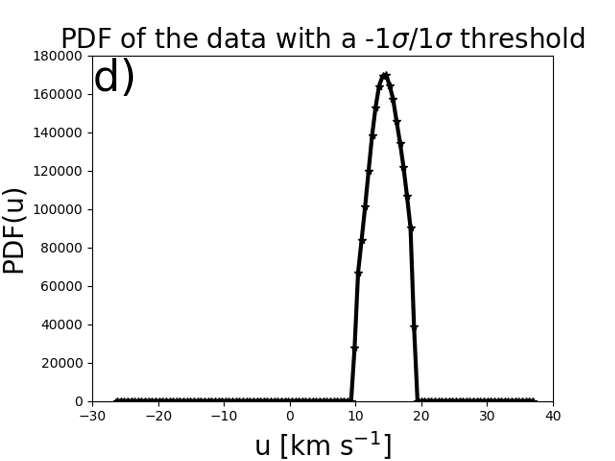}
	\caption[HAPDF]{PDF of the velocity field in H$\alpha$: a) PDF of the data filtered out using a $-9\sigma-5\sigma$ window centred in the mean of the raw data; b) PDF without the negative broad tail corresponding to a $-1\sigma/5\sigma$  filtering; c) PDF of the blue region corresponding to a filter of $-9\sigma/-1\sigma$; and d) PDF of the data filtered out to $-1\sigma/1\sigma$ of the mean of the raw data}
	\label{Fig:fig3}
\end{figure}

\begin{figure}
	\centering    
	\includegraphics[width=0.35\textwidth]{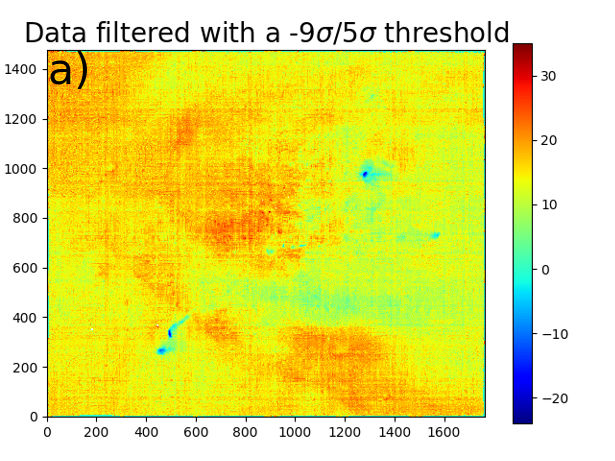}
	\includegraphics[width=0.35\textwidth]{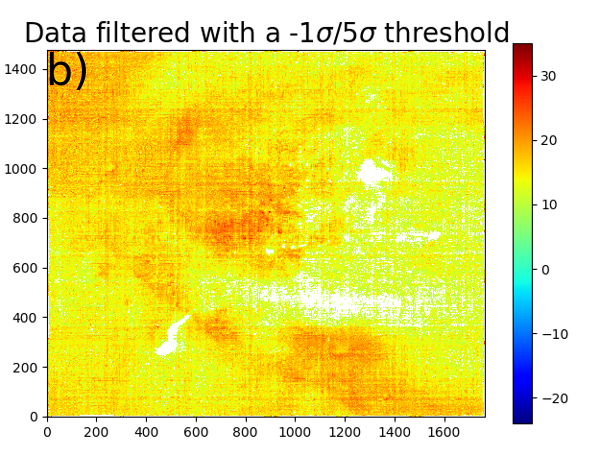}
	\includegraphics[width=0.35\textwidth]{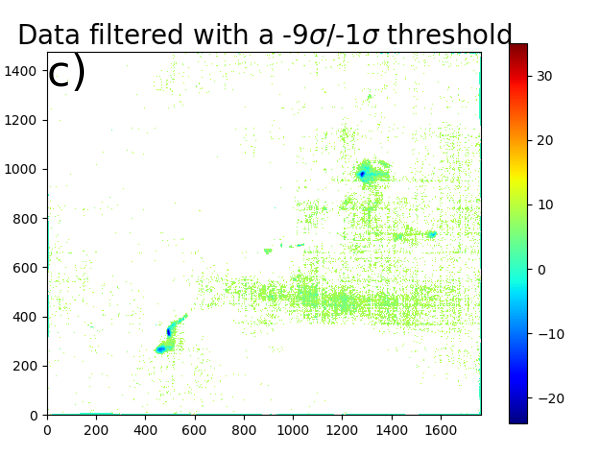}
	\includegraphics[width=0.35\textwidth]{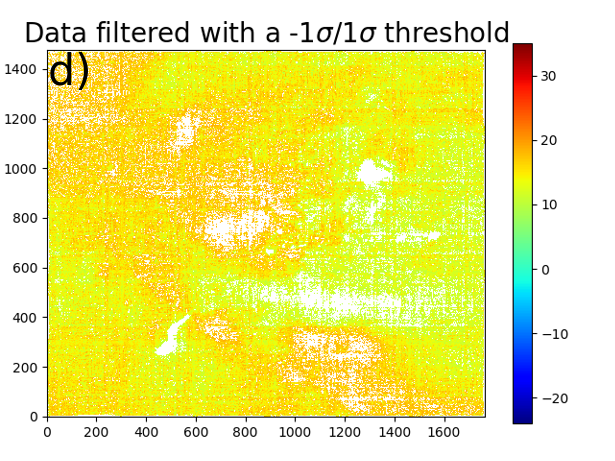}
	\caption[maps]{filtered H$\alpha$ maps: a) map filtered  using a $-9\sigma-5\sigma$ threshold; b) map filtered using a $-1\sigma/5\sigma$ threshold; c) map corresponding to the blue zone ($-9\sigma/-1\sigma$ filtering); and d) map corresponding to a $-1\sigma/1\sigma $ filtering.}
	\label{Fig:fig4}
\end{figure}

We filtered out the data such that only  velocities within -9$\sigma$/5$\sigma$ from the mean of the raw data are considered. This filters out the discontinuous  tails. We used this time 120 bins for visualization purposes. The resulting PDF is shown in Figure~\ref{Fig:fig3}a. The corresponding map, shown in Figure~\ref{Fig:fig4}a basically coincides with the map published by \citet{Weilbacher}. The zones marked out in their figure as the Big Arc East, the Big Arc South, the Bright Bar, the Red Bay and the Red Fan are all present as well as the general morphology of the field. However, now we have gained more insight about the PDF. This time, one can observe very clearly the central cusp between $u\approx 1-23$ km s$^{-1}$ followed by a heavy positive tail from $u\approx 23-37$ km s$^{-1}$. A secondary cusp between $u \approx -1-1$ km s$^{-1}$ followed by a heavy  tail from $u \approx -1-(-26)$ km s $^{-1}$ can also be observed. The mean of the filtered PDF is $\mu_u = 14.39$ km s$^{-1}$ and the standard deviation is $\sigma_u = 3.72$ km s$^{-1}$. The invariance of the mean implies that the raw-data long tails of up to $\pm$400 km s$^{-1}$  do not contribute significantly to the bulk of the PDF but their absence  decreases the standard deviation by $\sim 18$\%. 

\noindent{\bf Filter 2}

We noticed that negative and small positive velocities are present only on the light blue zones of the map, i.e. in the Big Arc South and close to the Herbig-Haro objects.  This might indicate that the gas contained there  is  uncoupled from the large-scale motion in the rest of the Huygens region. The motion in the Big Arc South may be indeed only associated to the Herbig-Halo objects and several bright stars. This is important to know at the time of calculating structure functions because it would not make sense to calculate a single structure function for two regions with uncoupled dynamics, i.e. in the case of using only the raw data or the data filtered out to -9$\sigma$/5$\sigma$. Because of this we filtered the PPV image to -1$\sigma$/5$\sigma$ which completely removes the singular negative tail of the distribution presented in Figure~\ref{Fig:fig3}a. The resulting distribution is presented in Figure~\ref{Fig:fig3}b, its mean is $\mu_u=$15.11  km s$^{-1}$ and its standard deviation is $\sigma_u=$2.80 km s$^{-1}$. The mean increases because we masked the small and negative values corresponding to the blue zones leaving only larger velocities, which produce also a smaller standard deviation. The corresponding image is shown in Figure~\ref{Fig:fig4}b. One can see there that all the blue zones including the Big Arc South have been completely filtered out (white dots). 

\noindent{\bf Filter 3}

 We decide also to analyse separately the blue zones in the map since we have shown above that they correspond to the negative tail of the -9$\sigma$/5$\sigma$ distribution and seem to manifest an uncoupled motion. Such broad negative-end tail  is presented in Figure~\ref{Fig:fig3}c and has a mean of $\mu_u=6.65$  km s$^{-1}$ and a standard deviation of $\sigma_u=$3.70 km s$^{-1}$. The corresponding image is shown in Figure~\ref{Fig:fig4}c.

 \noindent{\bf Filter 4}

Finally, we filtered  the data up to -1$\sigma$/1$\sigma$ which yields a mean of $\mu_u=$14.53  km s$^{-1}$ and a standard deviation of $\sigma_u=$2.27  km s$^{-1}$. The resulting PDF extends from 9-19  km s$^{-1}$. In this case, many   pixels belonging to the Big Arc South, Big Fan, Big Bay and the Bright Bar  are masked (see Figure~\ref{Fig:fig4}d). In the next section, we will show through structure functions that the unmasked yellow and red regions are responsible of providing most of the energy for the turbulent motion, even when the turbulence is more violent in the zones that correspond to the broad positive tail of the -9$\sigma$/5$\sigma$ distribution. This positive broad tail corresponds to the Red Fan, the Red Bay, an important  part of the Bright Bar and an  important part of the large region in the Northwest. We regard these  -1$\sigma$/1$\sigma$ masked field as a "mean field" that represents the background velocity field once the most prominent features have been eliminated.

\section{Structure functions}\label{SF}

A nebular analysis of the central part of the Orion Nebula (the Huygens region) was carried out by~\citet{McLeod} using the same {\emph MUSE} integral-field observations than here. They also attempted to diagnose the turbulence in the region using second order structure functions. For the latter, they used the Monte Carlo algorithm of~\citet{Boneberg} using only a sample of 10$^3$ pixels  whereas the data consists of $>2\times 10^6$ pixels.

Here, we are particularly interested in the structure of the projected velocity field obtained through the H$\alpha$ emission line since it allows to diagnose  turbulent motions from 0.00043-pc (the length covered by one pixel) up to 1-pc. The area of the central region of the Nebula is roughly 0.76$\times$0.63-pc$^2$, but we perform an analysis up to $\delta r\approx 1$-pc to probe the convergence of the $S_{\rm 2}$'s. For calculating the structure function we use the full data, i.e. we use every pixel of the map. We calculate the $S_2$'s using both a real space algorithm and the standard Fourier space algorithm.

\begin{figure}
	\centering    
	\includegraphics[width=0.5\textwidth]{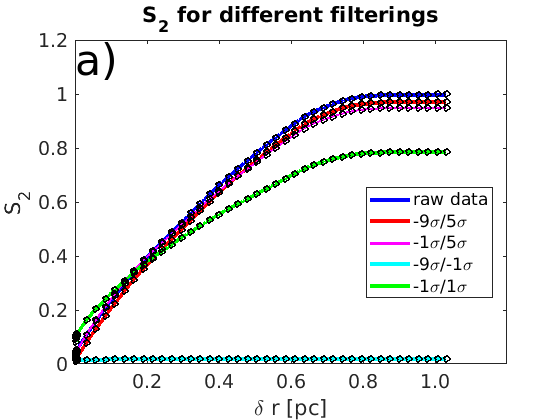}
	\includegraphics[width=0.5\textwidth]{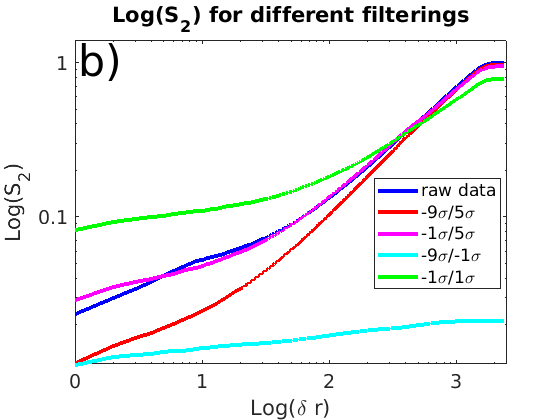}
	\caption[HAPDF]{Normalised S$_2$'s of the raw and filtered velocity data in H$\alpha$. The solid lines correspond to $S_2$'s computed by Fourier space methods. The black circles correspond to $S_2$'s calculated using a real space algorithm. a) $S_2$ in linear scale with the x-axis in pc b) $S_2$ in logarithmic scale with the x-axis in pixels}
	\label{Fig:fig5}
\end{figure}

On Figure~\ref{Fig:fig5}a, we present the normalised second order structure functions of the raw data and the filtered datasets. We can see that the $S_2$'s corresponding to the raw data, the $-9\sigma/5\sigma$ filtered data and the $-1\sigma/5\sigma$ filtered data look similar for $\delta r>300$ pixels. This occurs because the corresponding datasets have very similar (close to 100\%) kinetic energies per unit mass, $k$. The percentage of $k$ is obtained by multiplying the asymptotic value of each normalised $S_2$ by 100.  In Table~\ref{Tab:table1}, we give the percent of kinetic energy per unit mass that corresponds to each curve. We used the raw data as a reference. One can see that the $-1\sigma/1\sigma$ filtered data (green line) contains $\sim 79\%$ of $k$. This shows that the regions where the most violent turbulent motion occurs (the Red Fan, the Red Bay, the Bright Bar and a significant part of the large region to the Northwest of the image) contribute only with $\sim 16\%$ of $k$. These regions correspond statistically to the broad positive tail of the PDF.  Despite being large, the extended blue zone contributes only with $\sim$2\% of $k$.  The discontinuous long tails contribute with $\sim$3\% of $k$.

With exception of the $-9\sigma/-1\sigma$ dataset, all the curves have  initial segments that do not correspond to power laws (see Figure~\ref{Fig:fig5}b). Furthermore, a power law can be fitted from $\delta r=1$ to $\delta r=1500$ pixels  only to the  $-9\sigma/-1\sigma$ dataset. Nevertheless, we found that from $\delta r=300$ pixels (0.129-pc) to $\delta r=1500$ pixels\footnote{which is approximately the size of the North-South side of the data-box}  (0.645-pc) power laws of the form $ax^b$ describe well the shape of the other $S_2$'s. The amplitude and exponents of these power-laws are given in Table~\ref{Tab:table1}.

\begin{table}
	\centering
	\caption{Power laws to which the data was fitted. Their intervals of validity are given in pixels; one just needs to multiply by 0.00043 to convert to pc. The percent of the total energy corresponding to each dataset is also given. }
	\label{Tab:table1}
	\begin{tabular}{lcccr} 
		\hline
		Dataset & Fitted power-law & interval of validity&$R^2$& \% energy\\
		\hline
		raw &  $0.004656\times\delta r^{0.7287}$& 300-1500 &0.9997& 100.00\%\\
		-9$\sigma$/5$\sigma$ & $0.003338\times\delta r^{0.7668}$&300-1500 &0.9995 &97.30 \% \\
		-1$\sigma$/5$\sigma$ & $0.006191\times\delta r^{0.6824}$ &300-1500& 0.9998& 95.16\% \\
		-9$\sigma$/-1$\sigma$ & $0.01172\times\delta r^{0.08526}$ & 1-1500&0.9818& 2.14\% \\
		-1$\sigma$/1$\sigma$ &   $0.01271\times\delta r^{0.5566}$  &300-1500& 0.9996& 78.68\% \\
		\hline
	\end{tabular}
\end{table}

\section{Discussion}\label{discussion}

As considered by ~\citet{McLeod}, the turbulence in the Huygens region generates a velocity field characterised by stochastic hierarchical fluctuations that can be analysed through second-order structure functions. They calculated $S_2(\delta r)$ in real space using the Monte Carlo algorithm of~\citet{Boneberg}. Unfortunately, because of the random sampling characteristic of the algorithm, their calculation was incomplete and erroneous since they only used a sample   of $10^3$ pixels for calculating $S_2$, while the {\emph MUSE} dataset  consists of more than $2.6 \times 10^6$ pixels. Because of the above, for the H$\alpha$ velocity map, they obtained an exponent of $b=0.41$--$0.45$ for their computed $S_2$.

In turn, we found an exponent $b\sim 0.73$ for the raw data and of $b\sim 0.77$ for the $-9\sigma$/$5\sigma$ filtered data. The fact that the former exponent is smaller than the latter is due to the fact that the former has associated a larger amplitude $a$ such that the respective $S_2$  is always larger than the latter for  any given $\delta r$. Both exponents indicate a turbulence more violent than the  isotropic homogeneous turbulence defined by Kolmogorov, which has $b_{\rm K}=2/3$,~\citet{Kolmogorov1}. Nevertheless, by the evidence provided by the PDF analysis we performed and also by visual analysis,  the blue zones in the  map seem to be uncoupled from the dynamics of the rest of the map. Statistically, the blue zones correspond to the broad negative tail of the PDF and it includes a small secondary peak. If such a tail is removed, as in the case of the $-1\sigma/5\sigma$ filtered data the blue zones would not participate in the calculation of $S_2$ and one  would obtain an exponent $b\sim 0.68$, which is only slightly larger than  the exponent 2/3 predicted by the Kolmogorov theory.

The previous contradicts the conclusion of ~\citet{McLeod} which blamed the spectral and spatial resolution of {\emph MUSE } for the small exponents they obtained. They also pointed out that because of that, they were not able to reproduce the structure functions of previous works. However, two things must be remarked: a) first, they did not obtained the right exponents because of the Monte Carlo algorithm they used not because of the {\emph MUSE} resolution and b) second, the exponents obtained in the previous works  they cited  are flawed since they come from $S_2$'s that were evidently miscalculated, e.g. they do not display an asymptotic limit for large $\delta r$. So, we do not know neither the correct exponents for the data of previous works because of miscalculations. We suggest that a re-calculation and a re-analysis of the  data of previous works would be useful.

In our case, we find that when the negative broad tail of the PDF is ignored (the blue zone) and the positive tail is conserved, the yellow and red zones seem to exhibit an $S_2$ similar to that predicted by Kolmogorov for incompressible, homogeneous isotropic turbulence. This seems to indicate that the effect of compressibility is small in this case. Notice that the Red Fan, the Red Bay, the Bright Bar and the large zone at the Northwest of the image are prominently visible only when the positive broad tail is taken into account, compare Figures~\ref{Fig:fig4}a and \ref{Fig:fig4}b with Figure \ref{Fig:fig4}d. The latter Figure, which corresponds to the $-1\sigma/1\sigma$ filtering, can be thought as the representation of a "mean field" without the clustering structure produced by high velocity regions (red zones). These high velocity clusters correspond statistically to the positive tail of the PDF and they are responsible of making  the exponent of the $-1\sigma/5\sigma$ filtered data (Figure~\ref{Fig:fig4}b) almost coincide with the exponent of Kolmogorov turbulence. If the positive tail is also ignored we obtain the "mean field" which has an exponent $b=0.5566$ which is below the exponent of Kolmogorov turbulence.

By using a Monte Carlo algorithm, \citet{McLeod} did skip the contribution of the positive broad tail and thus obtained something similar to our data with a $-1\sigma/1\sigma$ filtering, which has an exponent $b$ below 2/3. The difference is that their 'unintentional' filtering due to the Monte Carlo algorithm they used was  more drastic than our $-1\sigma/1\sigma$ "mean field" filtering. That is the reason why sometimes the results from Monte Carlo algorithms are not reliable: they can induce a very drastic "mean field" filtering.

Finally, we remark that the blue zones enhance the contrast between velocities and because of that we obtain sharper $S_2$'s, however this is an effect that we do not need to consider if the dynamics of the blue zones is really detached from the dynamics of the rest of the gas. The blue zone per se has an $S_2$ which can be described by a power law of the form $ax^b$ from $\delta r=1$ pixel to $\delta r=1500$ pixels. However the exponent corresponding to such region is very small, $b=0.0861$ and the region contains only 2.18\% of the kinetic energy per unit mass.

\section{Conclusions}

We have calculated the second order structure functions of the Huygens Region of M 42 taking into account the velocity PDF of the PPV cubes we analysed. We found that when all the data in the map is considered, one can forecast a turbulence more violent than Kolmogorov turbulence since the slope of the inertial range is $0.72$--$0.77$. On the other hand, we found that if we ignore the zones that correspond to the broad negative tail of the velocity PDF we obtain an exponent of $0.6824$, which is close to 2/3, the exponent for Kolmogorov turbulence.  So, according to our diagnostic the gas exhibits a motion between Kolmogorov-type turbulence ($b=b_K=2/3$) and Burgers turbulence ($b=1$), see~\citet{Bouchaud} and \citet{Federrath2013}. Furthermore, the structure function of the raw data and the $-9\sigma/5\sigma$ filtered data are very close to the observational Larson law which generalises to $S_2\sim\delta r^{0.74-0.76}$. The previous law  have been found in Kolmogorov-Burgers turbulence where the dissipative structures are quasi-one dimensional shocks, see \citet{Boldyrev} and references therein.  Nevertheless, in our case we do not have a Kolmogorov-Burges regime but two separated regimes: an almost Kolmogorov regime and a secondary regime with a small exponent for the inertial regime and where bright stars and Herbig-Haro objects are present (the blue zone on Figure~\ref{Fig:fig1}).

The analysis of the velocity map by segmentation  using the velocity PDF is one of the differences of our work with previous ones. Another difference is that we obtained the  second structure functions both in Fourier and real space paying attention to every detail important for their calculation. Because of that our results are robust and reliable.

Finally, we suggest that the velocities in the negative tail of the PDF correspond to an extended region of ongoing star formation that has undergone local gravitational collapse, see~\citet{KlessenMC}. The presence of Herbig-Haro objects only in this extended region as well as the presence of bright stars in
the Big Arc South support this hypothesis.

\section{Acknowledgements}

 We thank Dr. Peter Weilbacher for kindly providing the PPV cubes that made this study possible. We also thank him for giving permission to reproduce an image  (Figure 1) from \citet{Weilbacher}. We also thank our anonymous referee whose recommendations helped to greatly improve this paper.








\bsp	
\label{lastpage}
\end{document}